\documentclass[conference]{IEEEtran}

\usepackage{graphicx}
\usepackage{amsmath,amssymb,amsfonts}
\usepackage{algorithm}
\usepackage{algorithmic}
\usepackage{psfrag}
\usepackage{subfigure}
\usepackage{url}
\usepackage{color}
\usepackage{soul}
\usepackage{xcolor}
\usepackage[]{hyperref}
\usepackage{lipsum}
\usepackage{array}
\usepackage{booktabs}
\newcommand{\head}[1]{\textnormal{\textbf{#1}}}
\newcommand{\normal}[1]{\multicolumn{1}{l}{#1}}

\graphicspath{{Figures/}}
\ifCLASSINFOpdf
\else
\fi

\begin{document}

\title{DeepWalking: Enabling Smartphone-based Walking Speed Estimation Using Deep Learning}

\author{Aawesh Shrestha and Myounggyu Won\\
WENS Lab, South Dakota State University, Brookings, SD, United States\\
\{aawesh.shrestha, myounggyu.won\}@sdstate.edu}%

\markboth{Journal of \LaTeX\ Class Files,~Vol.~13, No.~9, September~2014}%
{Shell \MakeLowercase{\textit{et al.}}: Bare Demo of IEEEtran.cls for Journals}

\maketitle

\begin{abstract}
Walking speed estimation is an essential component of mobile apps in various fields such as fitness, transportation, navigation, and health-care. Most existing solutions are focused on specialized medical applications that utilize body-worn motion sensors. These approaches do not serve effectively the general use case of numerous apps where the user holding a smartphone tries to find his or her walking speed solely based on smartphone sensors. However, existing smartphone-based approaches fail to provide acceptable precision for walking speed estimation. This leads to a question: is it possible to achieve comparable speed estimation accuracy using a smartphone over wearable sensor based obtrusive solutions?

We find the answer from advanced neural networks. In this paper, we present $\mathsf{DeepWalking}$, the first deep learning-based walking speed estimation scheme for smartphone. A deep convolutional neural network (DCNN) is applied to automatically identify and extract the most effective features from the accelerometer and gyroscope data of smartphone and to train the network model for accurate speed estimation. Experiments are performed with 10 participants using a treadmill. The average root-mean-squared-error (RMSE) of estimated walking speed is 0.16m/s which is comparable to the results obtained by state-of-the-art approaches based on a number of body-worn sensors (\emph{i.e.,} RMSE of 0.11m/s). The results indicate that a smartphone can be a strong tool for walking speed estimation if the sensor data are effectively calibrated and supported by advanced deep learning techniques.
\end{abstract}

\begin{IEEEkeywords}
Smartphone App,
Walking Speed Estimation
Deep Learning,
Convolutional Neural Network.
\end{IEEEkeywords}

\IEEEpeerreviewmaketitle

\section{Introduction}
\label{sec:introduction}

Walking speed estimation is an essential component for numerous smartphone apps in various domains. Accurate walking speed estimation is increasingly important for many health and fitness apps as usage of these apps has ever increased recently, over 300\% in just three years~\cite{fitness}. Research shows that walking speed information offers more than just the speed. In fact, walking speed can be used as the ``human odometer'' that allows us to derive valuable knowledge about the health status~\cite{hu2012model}. For example, changes in walking speed (monitored over a long term period) can be used as a vital clue to detect knee disabilities~\cite{andriacchi1977walking}. Recent studies suggest that walking speed is a useful indicator of the current lifestyle and future health predictor of a person~\cite{fritz2009white}. Other kinds of apps may also benefit significantly from precise walking speed estimation. For instance, accurate walking speed estimation is important for navigation apps such as Google Maps to provide reliable navigation results; transportation apps such as the ones designed for pedestrian safety~\cite{wang2012walksafe}\cite{jain2015lookup}\cite{won2018enabling} rely on walking speed estimation to reduce the number of pedestrian involved accidents.

Recent advances in mobile computing technologies~\cite{won2017hybridbaro} enabled numerous solutions have been proposed to predict walking speed effectively. A majority of these solutions are based on specialized body-mounted motion sensors~\cite{mannini2014walking}\cite{mcginnis2017machine}\cite{zihajehzadeh2016regression}. However, these systems are not adequate to support the general use cases of many mobile apps. There are smartphone-based solutions that utilize the embedded sensors of smartphone such as accelerometer~\cite{cox2014smartphone} and GPS~\cite{cho2010autogait}. However, imprecision of these inertial sensors results in unreliable walking speed estimation. Machine learning techniques are applied to enhance the speed estimation accuracy~\cite{park2012online}. However, a simple regression model and out-of-date machine learning techniques do not allow us to obtain highly comparable precision of walking speed estimation over wearable sensor based approaches, especially because it is challenging to identify and extract effective features.

With recent advances in deep learning techniques~\cite{lecun2015deep}, the parametrization of input data is no longer needed allowing us to perform highly accurate prediction at much faster speed with fully automated feature identification and extraction. In this paper, we present $\mathsf{DeepWalking}$, a deep learning based walking speed estimation framework for smartphone that can be easily integrated with various kinds of mobile apps. While deep learning has been used mostly for gait pattern recognition~\cite{gong2016deepmotion}\cite{gadaleta2018idnet}\cite{hannink2016stride}, to our knowledge, $\mathsf{DeepWalking}$ is the first smartphone based mobile system that maximizes precision of walking speed estimation using deep learning.

Especially, the deep convolutional neural network (DCNN) is known as a powerful deep learning technique that has seen successful applications in diverse areas such as human activity detection~\cite{ordonez2016deep}\cite{li2016deep}, and image recognition~\cite{krizhevsky2012imagenet}. We employ the DCNN to train the network model for walking speed estimation. Accelerometer and gyroscope sensor data collected with smartphone are fed into the DCNN, and appropriate features are automatically identified and extracted to build the network model. More precisely, noise of sensor data is effectively removed using a low-pass filter designed based on a power spectrum analysis of sensor data. Smartphone orientation independent vertical and horizontal components are extracted from the noise filtered accelerometer and gyroscope data. These vertical and horizontal components are used as primary data sources to construct the images that are provided as input to the DCNN for training the network model. The architecture of the DCNN is designed such that it minimizes the root mean squared error (RMSE) of walking speed estimation.

A proof-of-concept $\mathsf{DeepWalking}$ is implemented using a off-the-shelf smartphone. It was used to collect sensor data, build the DCNN model, and predict walking speed. Experiments were conducted with 10 participants using a treadmill to obtain the smartphone sensor data as well as the ground truth information. The results show that the RMSE of walking speed estimation is 0.16m/s. It is interesting to find that the speed estimation accuracy of $\mathsf{DeepWalking}$ quite comparable with the best result obtained with several skin-mounted accelerometer sensors. These findings indicate that a smartphone itself can actually be a powerful tool for walking speed estimation if sensor data are appropriately calibrated and supported by advanced machine learning techniques like DCNN.

The contributions of this paper are summarized as follows.

\begin{itemize}
\item To our knowledge, $\mathsf{DeepWalking}$ is the first deep learning based walking speed estimation framework specifically designed for smartphone apps.
\item $\mathsf{DeepWalking}$ is designed to serve general use cases and thus can be integrated with any app that requires accurate walking speed estimation.
\item A prototype of $\mathsf{DeepWalking}$ is implemented. Experiments were conducted with 10 participants to validate the effectiveness of $\mathsf{DeepWalking}$.
\end{itemize}

This paper is organized as follows. In Section~\ref{sec:related_work}, we review the literature on walking speed estimation. We then present an overview of the proposed walking speed estimation framework followed by the details of each system component in Section~\ref{sec:system_design}. Experimental results are discussed in Section~\ref{sec:experimental_results} after describing the experimental set up in Section~\ref{sec:experimental_setup}. We then conclude in Section~\ref{sec:conclusion}.


\section{Related Work}
\label{sec:related_work}

The majority of existing approaches for walking speed estimation are based on body-worn motion sensors (\emph{e.g.,} mostly accelerometer and gyroscope sensors secured to body)~\cite{herren1999prediction}\cite{vathsangam2010toward}\cite{mcginnis2017machine}\cite{sabatini2016ambulatory}\cite{hu2012model}\cite{li2010walking}. Mannini and Sabatini used foot mounted sensors~\cite{mannini2014walking}; McGinnis \emph{et al.} utilized wearable accelerometer arrays mounted on several parts of body like shank, sacrum, and thigh~\cite{mcginnis2017machine}; and Zihajehzadeh and Park exploited wrist-worn sensors for walking speed estimation~\cite{zihajehzadeh2016regression}. These approaches, however, do not serve effectively the general use case of numerous apps where the user holding a smartphone tries to find his or her walking speed solely based on smartphone sensors. Thus, a research question that we ask here is if it would be possible to achieve comparable precision of walking speed estimation using only the inertial sensors of smartphone, and what would be the methodologies to accomplish this goal.

A number of smartphone-based systems have been developed~\cite{cox2014smartphone}\cite{park2012online}\cite{cho2010autogait}. Cox~\emph{et al.} proposed a simple solution that estimates the walking speed based on the integration of acceleration~\cite{cox2014smartphone}. Cho~\emph{et al.} proposed to calibrate opportunistically the inertial sensor-based speed estimation using GPS of smartphone when the user is walking outdoors~\cite{cho2010autogait}. Park~\emph{et al.} applied the regularized kernel methods on the collected accelerometer data to achieve higher accuracy of walking speed estimation~\cite{park2012online}. Although there have been efforts to utilize machine learning techniques to improve the walking speed estimation accuracy, identification of effective features and manual extraction of those features are very challenging. In contrast to other smartphone based approaches, we leverage automated extraction of the most effective features using the DCNN to maximize the walking speed estimation accuracy.

The deep learning technology has been increasingly used in many recent works~\cite{gong2016deepmotion}\cite{gadaleta2018idnet}\cite{hannink2016stride}. However, these deep learning based systems are focused on recognition of gait patterns rather than estimation of walking speed. Gong~\emph{et al.} proposed a DCNN to perform gait assessment for multiple sclerosis patients based on the spectral and temporal associations among sensor data collected with a number of inertial body sensors~\cite{gong2016deepmotion}. Gadaleta and Rossi adopted the DCNN to recognize a target user based on the way of their walking utilizing the accelerometer and gyroscope data of smartphone~\cite{gadaleta2018idnet}. Hannink~\emph{et al.} used the DCNN to estimate the stride length~\cite{hannink2016stride}. In line of this research direction that is built upon deep learning technology, we propose the first general-purpose deep learning based walking speed estimation framework for smartphone.


\section{System Design}
\label{sec:system_design}

In this section, an overview of $\mathsf{DeepWalking}$ is presented followed by the details of each system component of $\mathsf{DeepWalking}$.

\subsection{System Overview}
\label{sec:system_overview}

\begin{figure}[!htbp]
\centering
\includegraphics[width=.8\columnwidth]{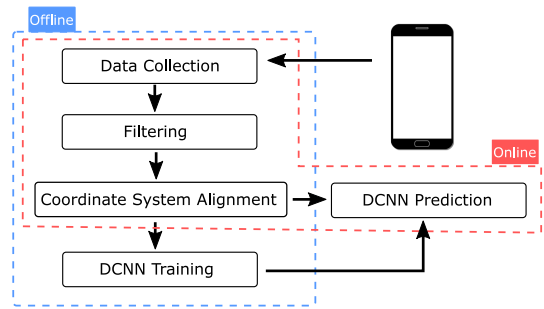}
\caption {System architecture of $\mathsf{DeepWalking}$ consisting of 5 key system components. The network model is trained in the offline mode, while prediction of walking speed is performed in the online mode.}
\label{fig:system_overview}
\end{figure}

Figure~\ref{fig:system_overview} displays the system architecture of $\mathsf{DeepWalking}$. $\mathsf{DeepWalking}$ consists of five key components: Data Collection, Filtering, Coordinate System Alignment, DCNN Training, and DCNN Prediction. The data collection module simply collects data from the accelerometer and gyroscope sensors of smartphone. The filtering module is designed to remove the noise of collected sensor data. And then, the coordinate system alignment module extracts phone orientation independent data from filtered sensor data. As shown in Figure~\ref{fig:system_overview}, these three system components are executed in both the online and offline modes. The network model for speed estimation is constructed in the offline mode by the DCNN Training module. This network model is used in the online mode to perform walking speed estimation by the DCNN Prediction module. More details on each system module are described in the following sections.

\subsection{Filtering}
\label{sec:denoising}

\begin{figure}[!htbp]
\centering
\includegraphics[width=.7\columnwidth]{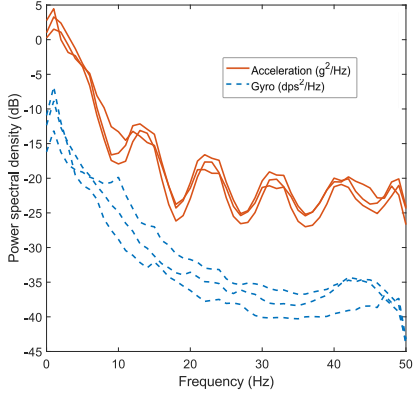}
\caption {Power spectral density of accelerometer and gyroscope data. Most of the power of acclerometer and gyroscope signals is concentrated at low frequencies.}
\label{fig:power_spec}
\end{figure}

To design a filter to remove noise of sensor data, we first investigate the power spectral density of the full accelerometer and gyroscope data that are used to train the DCNN network model. Figure~\ref{fig:power_spec} plots the power spectral density at different frequencies using the Welch's method with the Hanning window of 1 second and half window overlap~\cite{welch1967use}. The graph shows that most of the power for the accelerometer and gyroscope data are concentrated at low frequencies (\emph{i.e.,} between 0 and 15Hz). Accordingly, we designed a simple low-pass Finite Impulse Response (FIR) filter with a cutoff frequency of 15Hz. This filter is applied to sensor data to remove the noise.

\begin{figure}[!htbp]
\begin{minipage}[b]{0.48\columnwidth}
\centering
\includegraphics[width=\columnwidth]{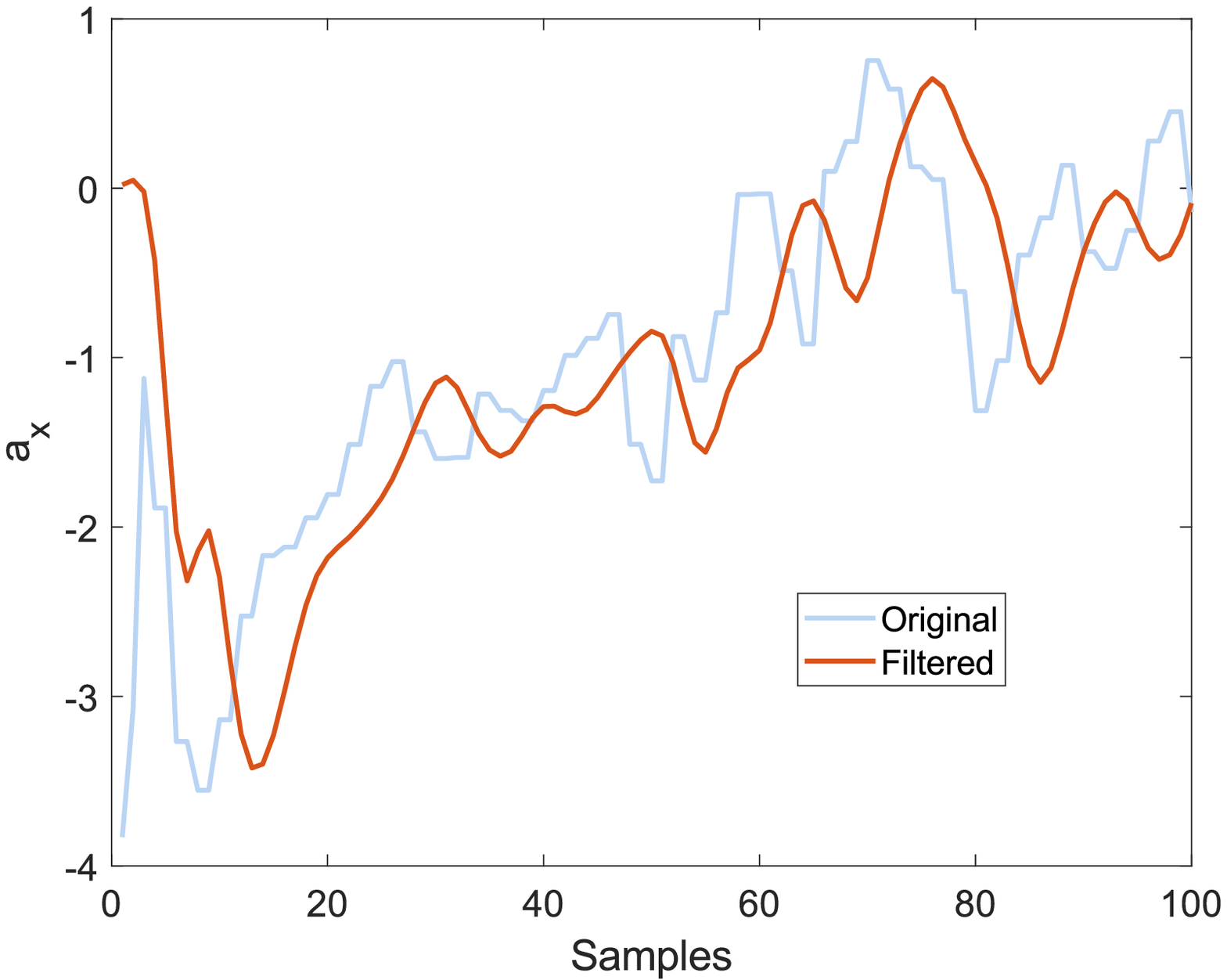}
\caption {Results of filtering for accelerometer data.}
\label{fig:denoising_acc}
\end{minipage}
\hspace{1mm}
\begin{minipage}[b]{0.48\columnwidth}
\centering
\includegraphics[width=\columnwidth]{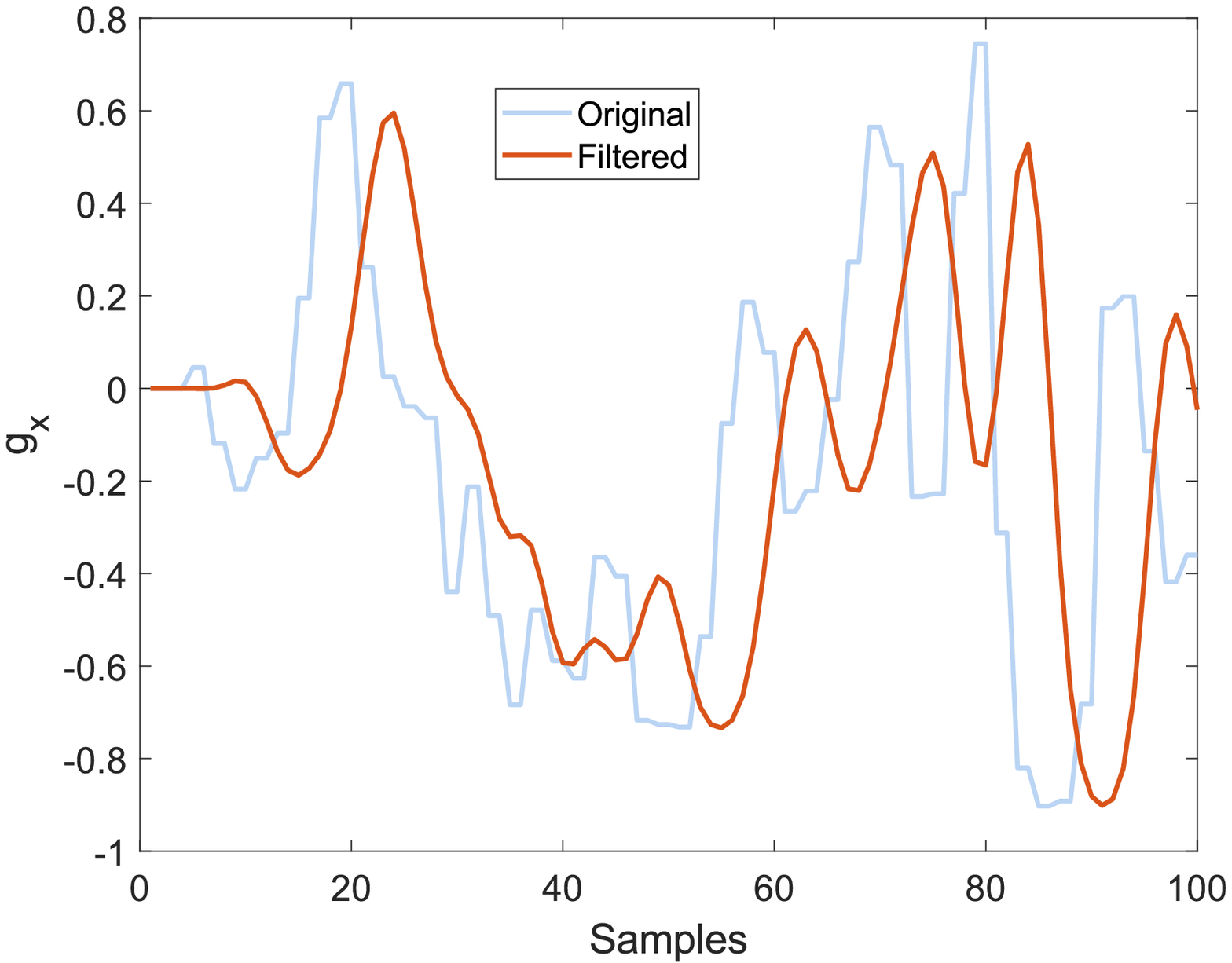}
\caption {Results of filtering for gyroscope data.}
\label{fig:denoising_gyro}
\end{minipage}
\end{figure}

Figures~\ref{fig:denoising_acc} and~\ref{fig:denoising_gyro} display the results of removing noise of accelerometer and gyroscope data using the low-pass filter respectively. As observed in these two figures, noise has been effectively reduced for both accelerometer and gyroscope data.

\subsection{Coordinate System Alignment}
\label{sec:orientation}

\begin{figure}[!htbp]
\centering
\includegraphics[width=.25\columnwidth]{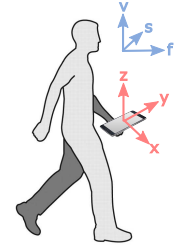}
\caption {The coordinate systems of smartphone and walking human. The varying orientation of smartphone is transformed into the fixed coordinate system of $(f,s,v)$.}
\label{fig:orientation}
\end{figure}

With the unknown orientation of smartphone, it is impossible to make reliable inference of walking speed given only accelerometer and gyroscope data. An algorithm is presented that transforms accelerometer and gyroscope data into orientation independent counterpart~\cite{mizell2003using}. Basically the algorithm is used to express accelerometer and gyroscope data on a fixed coordinate system. Specifically, Figure~\ref{fig:orientation} illustrates the three accelerometer/gyroscope axes denoted by $x$, $y$, and $z$. It also displays the fixed coordinate system defined for the user which is expressed with three axes $f$ (direction of walking motion), $v$ (vertical to the ground), and $s$ (orthogonal to the other two axes).

The coordinate system alignment module is started by deriving the gravity component on each axis (\emph{i.e.,} $x, y,$ and $z$) of the accelerometer. With the estimated gravity component, the magnitudes of vertical and horizontal components of accelerometer and gyroscope data are calculated. More specifically, let a vector $\mathbf{a}=\{a_x, a_y, a_z\}$ be a one-point acceleration measurement, where $a_x$, $a_y$, and $a_z$ represent acceleration measurements on respective axes. Assume that there are $N$ such acceleration vectors collected during a sampling interval. The gravity component denoted by a vector $\mathbf{v}=\{v_x,v_y,v_z\}$ is estimated by taking averages of all the measurements on each axis collected during the sampling interval, \emph{i.e.,} $v_x = \frac{\sum_{i=1}^{N}(a_x)}{N}$, $v_y = \frac{\sum_{i=1}^{N}(a_y)}{N}$, and $v_z = \frac{\sum_{i=1}^{N}(a_z)}{N}$. Note that we used 2 second as the sampling interval.

\begin{figure}[t]
\centering
\includegraphics[width=.8\columnwidth]{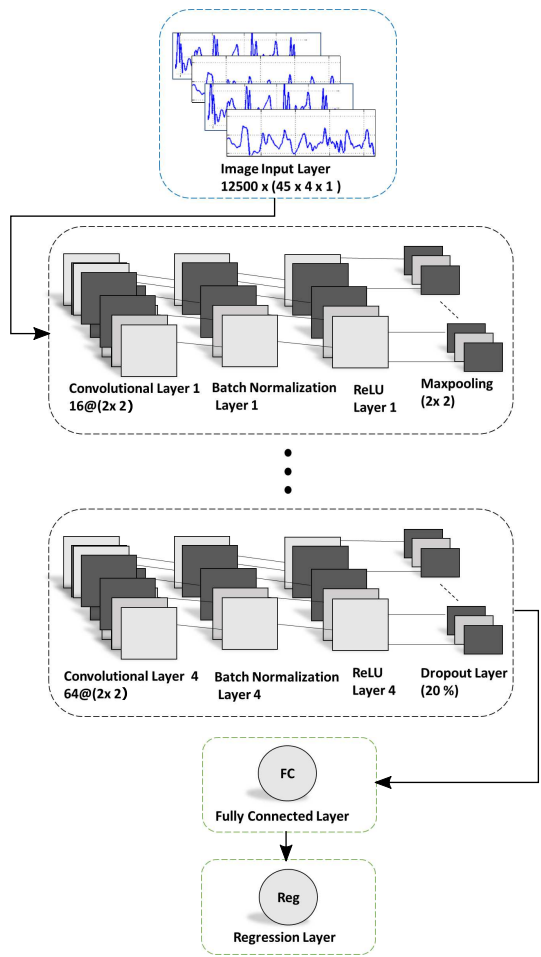}
\caption {The DCNN architecture. The architecture consists of cascaded four groups of layers (convolutional-batch normal-ReLU-Maxpooling) followed by the dropout layer, fully connected layer, and regression layer. }
\label{fig:dcnn}
\end{figure}

The dynamic component of $\mathbf{a}$ denoted by $\mathbf{d}$ that represents walking motion excluding the gravity component is $\mathbf{d}=\{a_x-v_x, a_y-v_y,a_z-v_z\}$. Now the vertical component $\mathbf{p}$ is calculated using projection of $\mathbf{d}$ onto the vertical axis $\mathbf{v}$ as follows.

\begin{equation}
\label{eq:1}
\mathbf{p}=\frac{\mathbf{d} \bullet \mathbf{v}}{\mathbf{v} \bullet \mathbf{v}}\mathbf{v}.
\end{equation}

Now the horizontal component $\mathbf{h}$ can be easily computed as $\mathbf{h} = \mathbf{d} - \mathbf{p}$. The vertical and horizontal components of gyroscope vector $\mathbf{g}=\{g_x,g_y,g_z\}$ are similarly calculated, \emph{i.e.,} the vertical component is calculated based on projection of $\mathbf{g}$ onto $\mathbf{v}$, and the horizontal component is computed by subtracting the vertical component from the gyroscope vector, \emph{i.e.,} $\mathbf{h} = \mathbf{g} - \mathbf{p}$. These vertical and horizontal components of accelerometer and gyroscope data are used as main sources to construct the images that are provided as input to the DCNN.

\subsection{Training and Prediction}
\label{sec:training}

Figure~\ref{fig:dcnn} illustrates the architecture of the DCNN. The input to the DCNN is $M$ images with the size of $45 \times 4 \times 1$. Each row of the image consists of 4 values that represent the magnitude of vertical and horizontal components of both accelerometer and gyroscope. We employ (16, 32, 48, 64) convolutional filters with the size of $2 \times 2$ and a stride of 1. This convolutional layer is a core building block of DCNN. It is applied to each input image by sliding the filters across the input vertically and horizontally and calculating the dot product of the weights of the filter and the input to optimize the weights.

We employ the batch normalization layer between the convolutional layer and the ReLU layer to lower the sensitivity to network initialization. In particular, we insert the ReLU layer to apply the non-saturating activation function $f(x)=\max(0,x)$ to increase the nonlinear properties of the network. Next, the maxpooling layer with the filter size $2 \times 2$ with a stride of 1 is inserted after the ReLU layer to reduce computation in the network and control overfitting. Four such groups of layers (\emph{i.e.,} convolutional layer, batch normalization layer, ReLU layer, and maxpooling layer) are cascaded with each other in a row--note that the maxpooling layer is excluded for the 4th group. We then add a dropout layer to alleviate the overfitting problem by randomly setting input elements to zero with a given probability--We found the dropout rate of 0.2 gave good results. Finally, the fully connected layer is used to compute the class scores which are then feed to the regression layer. More specifically, since our objective is to predict continuous data, walking speed, we add the regression layer at the end of the DCNN. In this process, root-mean-squared-error (RMSE) is used as the loss function.

\section{Experimental Setup}
\label{sec:experimental_setup}

We used Samsung Galaxy S6 to collect accelerometer and gyroscope data for both training the network model and prediction of walking speed. Ten volunteers participated in this data collection process. Each participant was asked to walk on a treadmill for 25 minutes, \emph{i.e.,} 5 minutes for each walking speed of $\{$1mph, 1.5mph, 2mph, 2.5mph, and 3mph$\}$. Consequently, we have obtained a total of 250 minutes of accelerometer and gyroscope data from 10 participants. The sampling frequency for both accelerometer and gyroscope sensors was set to 100Hz. During the process of data collection, the smartphone was placed in the participant's pocket.

To build the DCNN network model, 70\% of collected data (\emph{i.e.,} 17 minutes of sensor data for each participant) were used to create the images that were fed into the DCNN. The remaining 30\% of data (\emph{i.e.,} 8 minutes of sensor data for each participant) were used for prediction of walking speed. The prediction interval was 2 seconds, meaning that walking speed was estimated every 2 seconds.

The DCNN nework model was constructed offline using a PC that was equipped with the Intel core i5 6th generation CPU, 8GB of RAM running on Ubuntu 16.04 LTS. We used the neural network toolbox of MATLAB to create the network model. In addition, for proof-of-concept validation and more effective data analysis, prediction of walking speed was also performed under this offline configuration. However, it should be noted that the implemented system can be easily ported to any mobile platform like Android and iOS.

\section{Experimental Results}
\label{sec:experimental_results}

In this section, we evaluate the performance of $\mathsf{DeepWalking}$ focusing on measuring RMSE of walking speed estimation. Measured RMSE is compared with that obtained by state-of-the-art approaches based on body-worn motion sensors.

\subsection{Speed Estimation Accuracy}
\label{sec:accuracy}

\begin{figure}[!htbp]
\centering
\includegraphics[width=.8\columnwidth]{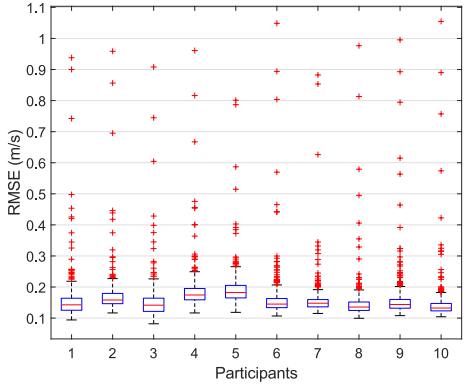}
\caption {RMSE of estimated walking speed for different participants. The RMSEs are comparable to state-of-the-art approaches based on wearable motion sensors.}
\label{fig:estimation_accuracy}
\end{figure}

We measured RMSE for 10 participants. Each participant was asked to walk on a treadmill for 5 minutes per each walking speed. The walking speed was estimated every 2 seconds (\emph{i.e.,} the prediction interval of 2 seconds). The results are displayed in Figure~\ref{fig:estimation_accuracy} as a box plot (the central mark: the median; the bottom and top edges of the box: the 25th and 75th percentiles, respectively; `+' symbol: outliers). As shown, we observe that overall the RMSE of all participants was quite small compared with the results of state-of-the-art work~\cite{mcginnis2017machine}. The average RMSE of all participants was 0.16m/s, and Table I shows the average RMSE for McGinnis \emph{et al.}~\cite{mcginnis2017machine} with varying sensor locations and numbers. The results indicate that a smartphone-based solution can be actually quite competitive in comparison with other approaches based on several skin-mounted sensors. More specifically, McGinnis \emph{et al.} achieved the best RMSE of 0.11m/s when the BioStampRC (a skin-mounted accelerometer sensor) is secured to the user's thigh, and shank~\cite{mcginnis2017machine}, while relatively lower RMSE was obtained when fewer accelerometers were used.

\begin{table}[!htbp]
    \centering
\begin{tabular}{@{}l*2{>{\ttfamily}l}%
  l<{\ttfamily}@{\ttfamily}}
  \toprule[1.5pt]
  & \multicolumn{2}{c}{\head{McGinnis \emph{et al.}}} &
    \multicolumn{1}{c}{\head{$\mathsf{DeepWalking}$}}\\
  & \normal{\head{Device Locations}} &
  \normal{\head{RMSE (m/s)}} & \normal{\head{RMSE (m/s)}}\\
  \cmidrule(lr){2-3}\cmidrule(r){4-4}
  &  Sacrum & 0.15 & \textbf{0.16} (Trouser Pocket)\\
  & Thigh & 0.15 & \\
  & Shank & 0.13 & \\
  & Sacrum, Thigh & 0.16 & \\
  & Sacrum, Shank & 0.13 & \\
  & Thigh, Shank & 0.11 & \\
  & Sacrum, Thigh, Shank & 0.12 & \\
  \bottomrule[1.5pt]
\end{tabular}
\caption{Comparison of RMSE of estimated walking speed under a treadmill setting (for healthy subjects).}
    \label{tab:rmse}
\end{table}

It is also observed that the largest RMSE gap between any two participants was only 0.04m/s. We interpret this result as an evidence that $\mathsf{DeepWalking}$ predicts walking speed reliably regardless of varying walking styles of different participants.

\subsection{Effect of Sensor Types}
\label{sec:effect_of_sensor_types}

Accelerometer and gyroscope are the most widely used motion sensors for walking speed estimation in the literature. In this section, we aim to understand the effect of varying combinations of these motion sensors on the walking speed estimation accuracy. More precisely, we generated different sets of images with varying types of sensors, \emph{i.e.,} ACC only, Gyro only, and ACC+Gyro. These images were individually fed into the DCNN to construct the network model and to predict walking speed.

\begin{figure}[!htbp]
\centering
\includegraphics[width=.8\columnwidth]{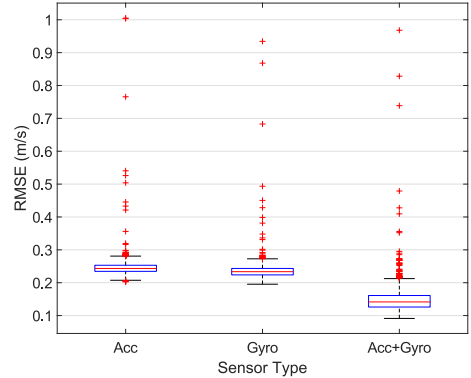}
\caption {Effect of sensor types on RMSE of estimated walking speed. When accelerometer and gyroscope sensor data are fusioned, RMSE was significantly reduced.}
\label{fig:effect_of_sensor}
\end{figure}

The results are shown in Figure~\ref{fig:effect_of_sensor}. As expected, we obtained the best accuracy when both accelerometer and gyroscope sensors were used together. Specifically, the RMSE of ACC+Gyro was 40.1\% and 37.5\% smaller compared with that of ACC-only and Gyro-only, respectively. An interesting observation was that the RMSE of ACC-only and Gyro-only was actually quite similar. This result indicates that both sensors are competitive in estimating walking speed, and when these two sensors are fusioned, the walking speed estimation accuracy can be improved significantly. Analyzing the effect of other embedded sensors of smartphone such as barometer, magnetometer, \emph{etc.} is left as future work.

\subsection{Effect of Number of Images}
\label{sec:effect_of_number_of_images}

Recall that $M$ images are fed into the DNN in order to construct the network model. In this section, we analyze the impact of the number of images $M$ on the walking speed estimation accuracy. For this experiment, we measured RMSE by varying $M$ from 1,000 to 12,500.

\begin{figure}[!htbp]
\centering
\includegraphics[width=.8\columnwidth]{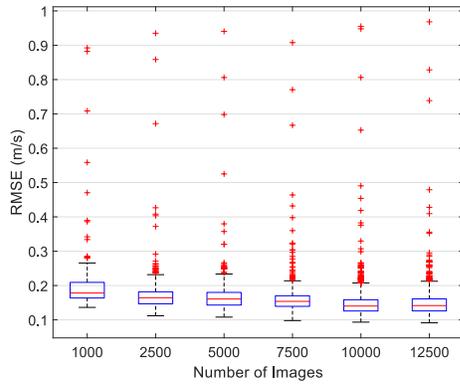}
\caption {Effect of number of images on RMSE of estimated walking speed. Larger the number of images, the smaller RMSE becomes. Yet, the performance gain also decreases as the number of images gets larger.}
\label{fig:effect_of_images}
\end{figure}

The results are depicted in Figure~\ref{fig:effect_of_images}. It was observed that as $M$ increased, the RMSE of estimated walking speed was gradually decreased. More specifically, when we increased $M$ from 1,000 to 12,500, the RMSE decreased by 29.4\%. The reason is that larger $M$ allows the DCNN to construct the network model that more effectively reflects the correlation between walking motion and walking speed.

Another interesting observation was that the performance gain (\emph{i.e.,} in terms of how much RMSE was decreased) decreased as $M$ was increased. These results indicate that while providing more input to the DCNN with larger $M$ is certainly beneficial, it does not necessarily lead to substantially improved performance when $M$ is sufficiently large enough. Now considering the overhead of training the network model with large $M$, it is important to decide an appropriate number of images that balances the tradeoff between the delay/overhead for training the network model and the expected performance gain. We leave this task of tuning the system parameter $M$ to balance the tradeoff as future work.

\section{Conclusion}
\label{sec:conclusion}

We have presented $\mathsf{DeepWalking}$, a deep learning based walking speed estimation framework for smartphone. $\mathsf{DeepWalking}$ is the first walking speed estimation system based on deep learning. It is expected to benefit diverse smartphone apps that rely on a system component of walking speed estimation.

Development of $\mathsf{DeepWalking}$ warrants a number of interesting future research directions. First, the prototype of $\mathsf{DeepWalking}$ will be fully implemented and tested with more diverse range of participants including different gender, age, and physical conditions and with different locations of the phone (\emph{e.g.,} in hand, on arm, \emph{etc.}). Second, case studies with fitness, transportation, and navigation apps will be performed to analyze the effect of accurate walking speed estimation that $\mathsf{DeepWalking}$ offers. Third, several system parameters (\emph{e.g.,} the number of images, different sensor types, hyper parameters for deep learning) will be tuned for better performance.

\section{Acknowledgement}
\label{sec:acknowledgement}

This research was supported in part by the Competitive Research Grant Program (CRGP) of South Dakota Board of Regents (SDBoR).

\bibliographystyle{IEEEtran}
\bibliography{mybibfile}



\end{document}